\documentclass{aa}
\usepackage{graphics}
\usepackage{txfonts}
\begin{document}


\title{Galactic interstellar $^{18}$O/$^{17}$O ratios - a radial gradient?
       \thanks{Figure~\ref{spectra} (the spectra) is available in electronic form}}

\author{J.G.A. Wouterloot
	  \inst{1} 
          \and
         C. Henkel
          \inst{2}
	  \and 
          J. Brand
	  \inst{3}
          \and
          G.R. Davis
          \inst{1}
	   }

\offprints{J.G.A. Wouterloot \email{j.wouterloot@jach.hawaii.edu}}

\institute{
Joint Astronomy Centre, 660 N. A'ohoku Place, Hilo, HI 96720, USA
\and Max-Planck-Institut f\"ur Radioastronomie,
Auf dem H\"ugel 69, 53121 Bonn, Germany \and
INAF - Istituto di Radioastronomia, Via Gobetti 101, 40129 Bologna, 
Italy}

\date{Received date/ Accepted date}

\abstract
{The determination of interstellar abundances is essential for a better 
understanding of stellar nucleosynthesis and the ``chemical'' evolution of 
the Galaxy.} 
{The aim is to determine $^{18}$O/$^{17}$O abundance ratios across the 
entire Galaxy. These provide a measure of the amount of enrichment 
by high-mass versus intermediate-mass stars.} 
{Such ratios, derived from the C$^{18}$O and C$^{17}$O $J$=1--0 lines 
alone, may be affected by systematic errors. Therefore, the C$^{18}$O 
and C$^{17}$O (1--0), (2--1), and (3--2), as well as the $^{13}$CO(1--0) 
and (2--1) lines, were observed towards 18 prominent galactic targets 
(a total of 25 positions). The combined dataset was analysed with a 
large velocity gradient model, accounting for optical depth 
effects.} 
{The data cover galactocentric radii between 0.1 and 16.9~kpc (solar 
circle at 8.5\,kpc). Near the centre of the Galaxy, $^{18}$O/$^{17}$O
= 2.88$\pm$0.11. For the galactic disc out to a galactocentric distance
of $\sim$10\,kpc, $^{18}$O/$^{17}$O = 4.16$\pm$0.09. At $\sim$16.5\,kpc
from the galactic centre, $^{18}$O/$^{17}$O = 5.03$\pm$0.46. Assuming
that $^{18}$O is synthesised predominantly in high-mass stars 
($M$$>$8\,M$_{\odot}$), while C$^{17}$O is mainly a product of lower 
mass stars, the ratio from the inner Galaxy indicates a dominance of 
CNO-hydrogen burning products that is also apparent in the carbon and 
nitrogen isotope ratios. The high $^{18}$O/$^{17}$O value of the solar 
system (5.5) relative to that of the ambient interstellar medium 
suggests contamination by nearby high-mass stars during its formation. 
The outer Galaxy poses a fundamental problem. High values in the 
metal-poor environment of the outer Galaxy are not matched by the low values 
observed towards the even more metal-poor Large Magellanic Cloud. 
Apparently, the outer Galaxy cannot be considered as an intermediate 
environment between the solar neighbourhood and the interstellar medium 
of small metal-poor galaxies. The apparent $^{18}$O/$^{17}$O gradient 
along the galactic disc and the discrepancy between outer disc and LMC 
isotope ratios may be explained by different ages of the respective 
stellar populations. More data from the central and far outer parts of 
the Galaxy are, however, needed to improve the statistical significance 
of our results.}
{}

\titlerunning{Interstellar $^{18}$O/$^{17}$O ratios}

\authorrunning{Wouterloot et al.}

\maketitle

%
%

\section{Introduction}

Isotope abundance ratios play an important role in our understanding 
of stellar nucleosynthesis and the secular ``chemical'' enrichment of 
the interstellar medium by stellar ejecta. While isotope ratios are 
not easily measured at optical wavelengths, observations of molecular 
clouds with their large number of molecular species allow us to 
distinguish between various isotopic species. In principle, accurate 
line intensity- and abundance-ratios can then be determined. 

A particularly useful tracer of nuclear processing and metal enrichment is 
the $^{18}$O/$^{17}$O ratio because $^{18}$O must be released primarily from 
high-mass stars ($M$$>$8M$_\odot$), while $^{17}$O may predominantly be 
ejected from stars with lower mass (Henkel \& Mauersberger \cite{henkelmau}; 
Henkel et al. \cite{henkelwil}; Langer \& Henkel \cite{langerhen}). Heger 
\& Langer (\cite{heger}) concluded that at the surface of rotating stars 
with masses of 8--25M$_\odot$, $^{17}$O is enriched and $^{18}$O is
depleted during hydrostatic burning prior to the supernova event. 
Most mass is ejected later, during the supernova explosion. While 
Hoffman 
et al. (\cite{hoffman}) conclude that production of both $^{18}$O and 
$^{17}$O in massive stars is reduced when taking into account new 
nuclear reaction rates, these reaction rates remain uncertain (see e.g. Stoesz 
\& Herwig \cite{stoesz}). To summarise, measured $^{18}$O/$^{17}$O ratios 
have the potential to provide relevant constraints for models of stellar 
nucleosynthesis and ``chemical'' evolution.

The $^{18}$O/$^{17}$O ratio is readily determined from the
C$^{18}$O/C$^{17}$O line intensity ratio. Both C$^{18}$O and C$^{17}$O have 
similar chemical and excitation properties. Rotational lines of both 
isotopologues are traditionally assumed to be optically thin, 
although C$^{18}$O can reach considerable optical depths in some regions 
(e.g., Bensch et al. \cite{bensch}). Oxygen isotopes are not 
fractionated (e.g., Langer et al. \cite{langergr}), which is related to 
the high first ionization potential of oxygen, 13.6~eV. Thus, O$^+$ 
should have a low abundance in molecular clouds and charge exchange reactions 
like those for carbon (see Eq.\,1 in Watson et al. \cite{watson}) should not 
significantly 
affect atomic and molecular oxygen abundance ratios. To summarise, observed 
CO line intensity ratios are a good measure of $^{18}$O/$^{17}$O.

Ratios of $^{18}$O/$^{17}$O have commonly been determined for clouds in the 
solar neighbourhood and the inner Galaxy (see the reviews of Wilson \& 
Matteucci 
\cite{wilmat}; Henkel et al. \cite{henkelwil}; Wilson \& Rood \cite{wilson}; 
Kahane \cite{kahane}; Bieging \cite{bieging}). For the galactic disc and 
centre region, Penzias (\cite{penzias}) reported average $^{18}$O/$^{17}$O 
ratios of 3.65$\pm$0.15 and 3.5$\pm$0.2, respectively. He found no 
significant gradient with galactocentric distance out to $R_{\rm GC}$=10~kpc 
($R_\odot$=8.5~kpc). The $^{18}$O/$^{17}$O ratio from HCO$^+$ $J$=1--0 in 
Sgr\,B2 (3.1$\pm$0.6; Gu\'elin et al. \cite{guelin}) is within the 
errors consistent with the CO results for the galactic centre region. 
The observed interstellar ratios as well as many ratios from  
envelopes of late-type stars of intermediate mass (some of them 
with $^{18}$O/$^{17}$O$<$1; e.g. Kahane et al. \cite{kahane92}) differ 
significantly from the solar system $^{18}$O/$^{17}$O ratio of 5.5, and 
Prantzos 
et al. (\cite{prantzos}) even considered the possibility of measurement 
errors in the galactic data.

The above-mentioned C$^{18}$O and C$^{17}$O observations were made in the 
$J$=1--0 transitions alone, which is not sufficient when trying to
account for radiative transfer effects. There exist also some $J$=2--1 
data (e.g. Hofner et al. \cite{hofner}; White \& Sandell \cite{white}), 
but these have been used to analyse C$^{18}$O optical depths and 
column densities, rather than studying the isotope ratio itself. Bensch 
et al. (\cite{bensch}) observed the definitely optically thin $^{13}$C$^{18}$O 
and $^{13}$C$^{17}$O $J$=1--0 lines and derived 4.15$\pm$0.52 at one 
position in the nearby ($d$$\sim$140\,pc) $\rho$ Oph cloud. The same cloud 
was observed towards 21 positions in the $J$=1--0, 2--1, and 3--2 transitions 
of C$^{18}$O and C$^{17}$O by Wouterloot et al. (\cite{wbh05}). Large Velocity
Gradient (LVG) model 
calculations combining these data resulted 
in a similar but more accurate value, 4.11$\pm$0.14, for the 
$^{18}$O/$^{17}$O ratio. Very recently, Zhang et al. (2007) measured the 
$J$=1--0 lines towards a selected area of the southern star forming region 
NGC\,6334 ($d$$\sim$1.7\,kpc). Under the (in this case realistic) 
assumption that all C$^{18}$O lines were optically thin, they 
obtained a ratio of 4.13$\pm$0.13.

To date, little has been done to analyse molecular abundances in clouds at 
the edge of the Galaxy (for a pioneering study, see Wouterloot \& 
Brand \cite{woutbrand}). With respect to the solar neighbourhood, the 
far-outer Galaxy, at galactocentric radii of $R_{\rm GC}$$>$16~kpc, is 
characterised by a low metallicity, a weak interstellar radiation field, 
a small cosmic ray flux, and a low large-scale average gas pressure  
(Brand \& Wouterloot \cite{brand}, Ruffle et al. \cite{ruffle}, and 
references therein).

Our initial measurements of two objects at large galactocentric 
radii, \object{DDT94 Cloud 1} and \object{DDT94 Cloud 2} (Digel et al. 
\cite{digel}), with distances of $R_{\rm GC}$ of 22 and 17$\pm$2~kpc 
(Smartt et al. \cite{smartt}) did not provide conclusive results. 
Therefore, to derive more reliable values, we have observed
four objects from the catalogue of Wouterloot \& Brand (\cite{wb89}) 
with stronger C$^{18}$O lines ($T_{\rm{A}}^*$ $\sim$ 1~K versus 0.2~K 
for the DDT94 clouds) to cover the range of $R_{\rm{GC}}$ beyond 
15~kpc. In addition, we have reobserved objects at 0~kpc $\la$ 
$R_{\rm{GC}}$ $\la$ 10~kpc to compare these results with previous 
estimates based on the $J$=1--0 line alone. Our sample consists of 
25 positions in 18 sources that are observed in the $J$ = 1--0, 2--1, 
and 3--2 rotational lines of C$^{18}$O and C$^{17}$O. $^{13}$CO(1--0) 
and (2--1) data were also taken.

%
%
\begin{table}
\caption[]{Observed transitions.
\label{freqs}}
\begin{flushleft}
\begin{tabular}{lrccc}
\hline\noalign{\smallskip}
Molecule & Frequency& HPBW$^a$ & $B_{\rm{eff}}^b$ & $F_{\rm{eff}}^b$ \\
 & (MHz) & (arcsec) &      &        \\
\hline\noalign{\smallskip}
$^{12}{\rm C}^{18}{\rm O}$(1$-$0)&109782.182&21&0.68&0.92 (1995,1996)  \\
                        &    &  & 0.76  &0.92 (05-2000) \\
                        &    &  & 0.76  &0.92 (09-2000) \\
$^{12}{\rm C}^{17}{\rm O}$(1$-$0)&112359.277&21&0.68&0.92 (1995,1996) \\
$^{13}{\rm C}{\rm O}$(1$-$0)     &110201.370&21& 0.76 &0.90 (05-2000)\\
                        &    &  & 0.76 &0.92 (09-2000)\\
$^{12}{\rm C}^{18}{\rm O}$(2$-$1)&219560.319&11&0.41&0.86 (1995,1996)   \\
                                 &   &  & 0.54  &0.91 (05-2000) \\
                                 &   &  & 0.54  &0.85 (09-2000) \\
$^{12}{\rm C}^{17}{\rm O}$(2$-$1)&224714.370&11&0.41&0.86 (1995,1996) \\
$^{13}{\rm C}{\rm O}$(2$-$1)     &220398.686&11& 0.54 & 0.92 (05-2000) \\
                           &    &  & 0.54 & 0.85 (09-2000) \\
$^{12}{\rm C}{^{18}\rm O}$(3$-$2) & 329330.545 & 14 & 0.71 & -  \\
$^{12}{\rm C}{^{17}\rm O}$(3$-$2) & 337061.129 & 14 & 0.71 & -  \\
\noalign{\smallskip}
\hline
\multicolumn{5}{l}{$^{a}$: Half Power Beam Width}\\
\multicolumn{5}{l}{$^{b}$: Beam ($B_{\rm eff}$) and forward hemisphere 
($F_{\rm eff}$) efficiency. To convert } \\
\multicolumn{5}{l}{$T_{\rm A}^*$ antenna temperatures to $T_{\rm mb}$ main
beam brightness temperatures,}\\
\multicolumn{5}{l}{multiply the $J$=1--0 and 2--1 values by 
$F_{\rm eff}$/$B_{\rm eff}$ (see, e.g.,}\\
\multicolumn{5}{l}{Rohlfs \& Wilson 1996). For the $J$=3--2
lines, multiply by 1/$B_{\rm eff}$.}\\
\end{tabular}
\end{flushleft}
\end{table}
%
%

\section{Observations}

\subsection{IRAM 30-m}

On March 4, 1995, we used the IRAM 30-m telescope\footnote{IRAM is 
supported by INSU/CNRS (France), the MPG (Germany), and the IGN (Spain).} 
to observe C$^{18}$O and C$^{17}$O (1--0, 2--1) towards DDT94 Cloud 1 
and 2. Between July 28 and August 2, 1995, we used the same telescope 
to observe C$^{18}$O (1--0, 2--1) and C$^{17}$O (1--0, 2--1) (see 
Table~\ref{freqs}) towards the other clouds of our sample. We used 
three SIS receivers simultaneously (first for C$^{18}$O $J$=1--0, 
2--1 and C$^{17}$O 2--1, then for C$^{17}$O 1--0, 2--1 and C$^{18}$O 
2--1; all data were consistent) in combination with an autocorrelator 
split into three parts with equal velocity resolutions (in most cases 
0.10~km$\,$s$^{-1}$). Towards most positions we initially used 
frequency-switching (by 7.9 and 15.8~MHz at 3~mm and 1.3~mm, respectively). 
Towards sources with broad lines, however, we observed in a position-switching 
(total power) mode. Offsets of the reference positions are given in 
Table~\ref{sources}. Velocity resolutions and rms values of the spectra 
are displayed in Table~3 (see Sect. 3). At positions with weak C$^{17}$O 
lines, frequency-switched spectra showed prohibitively large 
baseline ripples. Therefore these sources were also observed in total 
power mode. From most spectra obtained in total power mode we subtracted 
a low order baseline. Remaining baseline ripples due to standing waves 
were removed by subtracting a sinusoidal baseline. 

To investigate the importance of radiative transfer effects we observed
$^{13}$CO (1--0) and (2--1) simultaneously with C$^{18}$O (1--0) and (2--1) 
between May 10 and 15 and on September 7, 2000. All observations were made 
in a total power mode. For the four sources from the sample of WB89, the
$^{13}$CO and C$^{18}$O data
were taken from Wouterloot \& Brand (\cite{woutbrand}). 

On April 4-5, 2005, we used HERA to observe C$^{18}$O (2-1) towards all 
sources. HERA consists of two 9 pixel receivers on 3x3 arrays (a 
24$^{\prime\prime}$ raster). The spectral resolution was 0.11~km/s and 
observations were made using position switching. These data are used to 
convolve the $J$=2--1 and $J$=3--2 spectra to the IRAM $J$=1--0 angular 
resolution.

An rms pointing accuracy of about 3\arcsec\ was determined from scans across 
nearby continuum sources. Planet measurements showed that the receiver 
alignment was within 1--4\arcsec. The observed transitions, frequencies, 
beamsizes, beam and forward efficiencies are given in Table~\ref{freqs}. 
In the following, $T_{\rm A}^*$ intensities are given, unless specified 
otherwise.

\subsection{JCMT 15-m}

On July 14, 2001, we observed C$^{17}$O(3--2) towards W49N and W51d with the 
15-m James Clerk Maxwell Telescope (JCMT)\footnote{The JCMT is operated
by The Joint Astronomy Centre on behalf of the Science and Technology 
Facilities Council of the United Kingdom, the Netherlands Organisation for 
Scientific Research, and the National Research Council of Canada. Our
programs: m04bd10 and m05bi17. JCMT archive data were obtained from the 
Canadian Astronomy Data Centre, which is operated by the Herzberg Institute 
of Astrophysics, National Research Council of Canada.} using position 
switching. On various occasions between November 7, 2004, and May 19, 2005, 
we observed C$^{17}$O(3--2) towards NGC2024, L134N, and four offset positions 
near Ori-KL. C$^{18}$O(3--2) was observed towards the same sources and 
towards WB89-501. On February 10, 2006, we observed the remaining sources 
(except for the WB89 objects and G34.3+0.2, W3OH, and NGC7538) in C$^{18}$O 
and C$^{17}$O(3--2).

The JCMT observations were made with the 2x800 channel DAS autocorrelator 
and the dual polarization RxB3 SIS receiver using position switching. The 
channel spacing was 156 -- 625~kHz and the bandwidth was 250 -- 
920~MHz, depending on the line width of the source. Offsets of the reference 
positions used are given in Table~\ref{sources}. Pointing was done on nearby 
continuum sources and was generally accurate to within 3$^{\prime\prime}$.
 
From the JCMT archives we obtained C$^{17}$O(3--2) and C$^{18}$O(3--2) data
towards G34.3+0.2, W3OH, and NGC7538. The JCMT archival data are from standard 
sources which are frequently observed, and always after doing a pointing 
observation on the source itself. The coordinates are within a few arcsec 
(much less than the beamsize and within the pointing accuracy) of the 
positions used for our IRAM 30-m observations. In some cases several 
spectra were available. Discrepant results and measurements with too low
velocity resolution, affecting peak intensities, were not considered and 
the remaining spectra were averaged.

\subsection{Calibration uncertainties}

Calibration uncertainties in individual spectra may amount to $\pm$10\% 
which corresponds to $\pm$14\% in line intensity ratios. Changes in telescope
efficiencies (see Table~1) may also contribute to the error budget. Like the 
uncertainties in the efficiencies, pointing errors may also affect those 
intensity ratios where the two lines were measured separately. This holds for 
all JCMT observations. As mentioned in Sect.\,2.2, however, results from 
repeatedly 
observed sources turn out to be consistent. A direct way to assess the quality 
of the IRAM data is provided by Table~3. Here, line parameters for the 
C$^{18}$O $J$=1--0 and 2--1 lines are given twice, because the lines were 
once observed together with C$^{17}$O and another time with $^{13}$CO. 
The uncertainty in the IRAM line ratios will be less than those in the
line intensities which are suggested by differences between both sets of 
C$^{18}$O measurements, because the ratios are commonly
based on simultaneous observations. Nevertheless, the data displayed in 
Table~3 
provide firm upper limits to the full error budget of the IRAM line 
intensities. 

%
%
\begin{table*}
\caption[]{Observed sources$^{a}$.
\label{sources}}
\begin{flushleft}
\begin{tabular}{lccrrrrrrc}
\hline\noalign{\smallskip}
\multicolumn{1}{l}{Source}& \multicolumn{1}{c}{$\alpha$(2000)}& 
\multicolumn{1}{c}{$\delta$(2000)}& 
\multicolumn{1}{c}{off $\alpha$}& 
\multicolumn{1}{c}{off $\delta$}& \multicolumn{1}{c}{ref $\alpha$}& 
\multicolumn{1}{c}{ref $\delta$}& 
\multicolumn{1}{c}{$d$}& \multicolumn{1}{c}{$R_{\rm GC}$} &
\multicolumn{1}{c}{ref} \\
\multicolumn{1}{l}{}& \multicolumn{1}{c}{{\sl h\ m\ s}}&
\multicolumn{1}{c}{\degr\ \arcmin\ \arcsec}& \multicolumn{1}{r}{\arcsec}&
\multicolumn{1}{r}{\arcsec}& \multicolumn{1}{r}{\arcmin}&
\multicolumn{1}{r}{\arcmin}& 
\multicolumn{1}{c}{kpc}& \multicolumn{1}{c}{kpc} &  \\
\hline\noalign{\smallskip}
\object{WB89 380}&01 07 51.3&+65 21 25 &0 &0 &1.2 &16.7&10.4&16.8 & (1) \\
\object{WB89 391}&01 19 25.4&+65 45 50 &0 &0 & 1.2&16.7&10.4&16.9 & (1) \\
 & & & & &$-$60 & 0 & & \\
\object{DDT94 Cloud 1}& 02 04 51.9 & +63 07 41 & 0 & 0 & & & 16.4 & 22.9 & (2) \\
\object{W 3} &02 25 44.2 &+62 06 01 & 0 & 0 & & & 2.0 & 10.0 & (3) \\
\object{W 3OH} &02 27 04.2 &+61 52 25 &0&0 &$-$60&0& 2.0 & 10.0 & (3) \\
 & & & & & 120 & 0 & & & \\
\object{WB89 437}&02 43 29.0&+62 57 16 &0&0&$-$13.7&6.2&9.1&16.2 & (1) \\
 & & & & &54.8 & $-$24.8 & & & \\
\object{DDT94 Cloud 2}&02 48 30.4 &+58 27 47 &0&0& & & 10.0 &17.0 & (4) \\
\object{WB89 501}&03 52 27.7&+57 48 26 &0&0&$-$11&9&8.6&16.3 & (1) \\
 & & & & &12.2 & --10 & & & \\
\object{Ori KL} &05 35 14.2&$-$05 22 26 &0&0&$-$30&0 & 0.41 & 8.84 & (5) \\
       &          &           &  &       &45&0 & & & \\
       &          &           &30&$-$240 &    &  & & & \\
       &          &           &30& 30    &    &  & & & \\
       &          &           & 0&$-$150 &    &  & & & \\
\object{NGC 2024}&05 41 45.4&$-$01 54 47 & 0&  0&15&0 & 0.42 & 8.86 & (6) \\
        &          &           &  &     &30&0 & & & \\
        &          &           &  &     &$-$30&0 & & & \\
\object{L 134N}  &15 54 06.6&$-$02 52 19 & $-$84&$-$48&20&10& 0.11 & 8.41 & (7) \\
\object {Sgr B2(M)}&17 47 20.4&$-$28 23 02 &0&0&$-$30 &60 & 8.5& 0.1 & (8) \\
                 &  & & 0 &15&$-$30 &60 & 8.5& 0.1\\
\object{W 33} &18 14 14.4 &$-$17 55 50 &0&0&$-$30&30& 3.8&4.8 & (2)\\
 & & & 0 & 24 & & &  &\\
\object{G34.3+0.2} &18 53 18.4 &+01 14 56 &0&0&$-$30&30& 3.9&5.7 & (2) \\
\object{W 49N} &19 10 13.4 &+09 06 14 & 0 & 0 &0 &$-$60 & 11.4 & 7.8 & (9) \\
\object{W 51}d &19 23 39.6 &+14 31 13 & 0 & 0 &0 &$-$60 & 7.0 & 6.6 & (10)\\
\object{DR 21} &20 38 54.6 &+42 19 23 & 0 & 0 & & & 2.0&8.5 & (11)\\
DR~21cal &20 39 01.2 &+42 19 45 & 0 & 0 & & & 2.0&8.5 & (11) \\
\object{DR 21OH} &20 39 01.9 &+42 22 46 & 0 & 0 & & & 2.0&8.5 & (11)\\
\object{NGC 7538} &23 13 45.2 &+61 28 10 &0&0&0&$-$60& 2.7&9.8 & (12)\\
 & & & & &0 & 60 & & \\
\noalign{\smallskip}
\hline
\end{tabular}
\end{flushleft}
(1) Brand \& Wouterloot (\cite{bw1994}); peak position from Wouterloot \& Brand 
     (\cite{woutbrand}).\\
(2) Kinematic distance from the observed radial velocity and the rotation curve of 
     Brand \& Blitz (\cite{bb93}). For W33 and G34.3+0.2, the near kinematic distance 
     is given. \\
(3) Weighted average of distances of Hachisuka et al. (\cite{hachisuka}), 2.04$\pm$0.07~kpc, 
    Xu et al. (\cite{xu}), 1.95$\pm$0.04~kpc, and Imai et al. (\cite{imai}), 1.83$\pm$0.14~kpc. \\
(4) Smartt et al. (\cite{smartt}). \\
(5) Menten et al. (\cite{menten}). \\
(6) Anthony-Twarog (\cite{anth-twarog}). \\
(7) Franco (\cite{franco}). \\
(8) Kerr \& Lynden-Bell (\cite{kerr}); we note that Eisenhauer et al. (\cite{eisenhauer}) derived
    7.62$\pm$0.32\,kpc for the distance to the Galactic Centre which may be compared with the 
    IAU value of 8.5 kpc assumed in this work. \\
(9) Gwinn et al.(\cite{gwinn}). \\
(10) Lacy et al. (\cite{lacy}). \\
(11) Dickel et al. (\cite{dickel}). DR 21cal position from Mauersberger et al. (\cite{mauersberger}).\\
(12) From trigonometric parallax of CH$_3$OH masers, L. Moscadelli 
     pers. comm. \\
$^{a}$: Source names (Col.\,1); Source positions in equatorial coordinates 
(Cols.\,2 and 3); Observed offset positions (Cols.\,4 and 5); 
Offsets of the\\ 
reference positions used for total power measurements 
(Cols.\,6 and 7); Adopted distances from the Sun
and the galactic centre (Cols.\,8 and 9);\\
Reference to these values (Col. 10). \\
\end{table*}
%
%

\section{Results}

Table~\ref{sources} contains the sample of observed sources. Sgr\,B2 
is the most prominent star-forming region close to the galactic centre, 
L\,134N is a dark cloud near the Sun, and WB89\,380, WB89\,391, WB89\,\,437, 
and WB89\,501 are sources located in the far outer Galaxy. The other targets
are star forming regions located at an intermediate range of galactocentric 
radii, extending from the inner ``molecular ring'' out to the Perseus arm. 

The complete set of spectra is displayed in the appendix (Fig.~\ref{spectra}) 
that is available in electronic form. All spectra of an individual source are 
plotted on the same velocity scale. Line parameters obtained from Gaussian 
fits to the $J$=1--0, 2--1, and 3--2 transitions are summarised in 
Table~\ref{results}. To compare abundances and not line intensities, 
we corrected the C$^{18}$O/C$^{17}$O line intensity ratios for the difference 
in frequency of both isotopologues, following Linke et al. (\cite{linke}) and 
Penzias (\cite{penzias}). This implies an increase of the ratios by a factor 
1.047. For W51d, all parameters were calculated for two velocity intervals, 
accounting for the two emission lines at +50 and +60\,km\,s$^{-1}$ detected 
towards this source. The new C$^{18}$O HERA line parameters towards the 
WB89 sources agree with those obtained earlier, also with the 30-m telescope, 
by Wouterloot \& Brand (\cite{woutbrand}).

\begin{table*}[t]
\label{results}
\vspace{3.8cm}
\hspace{-1.2cm}
\resizebox{13.5cm}{!}{\rotatebox{180}{\includegraphics{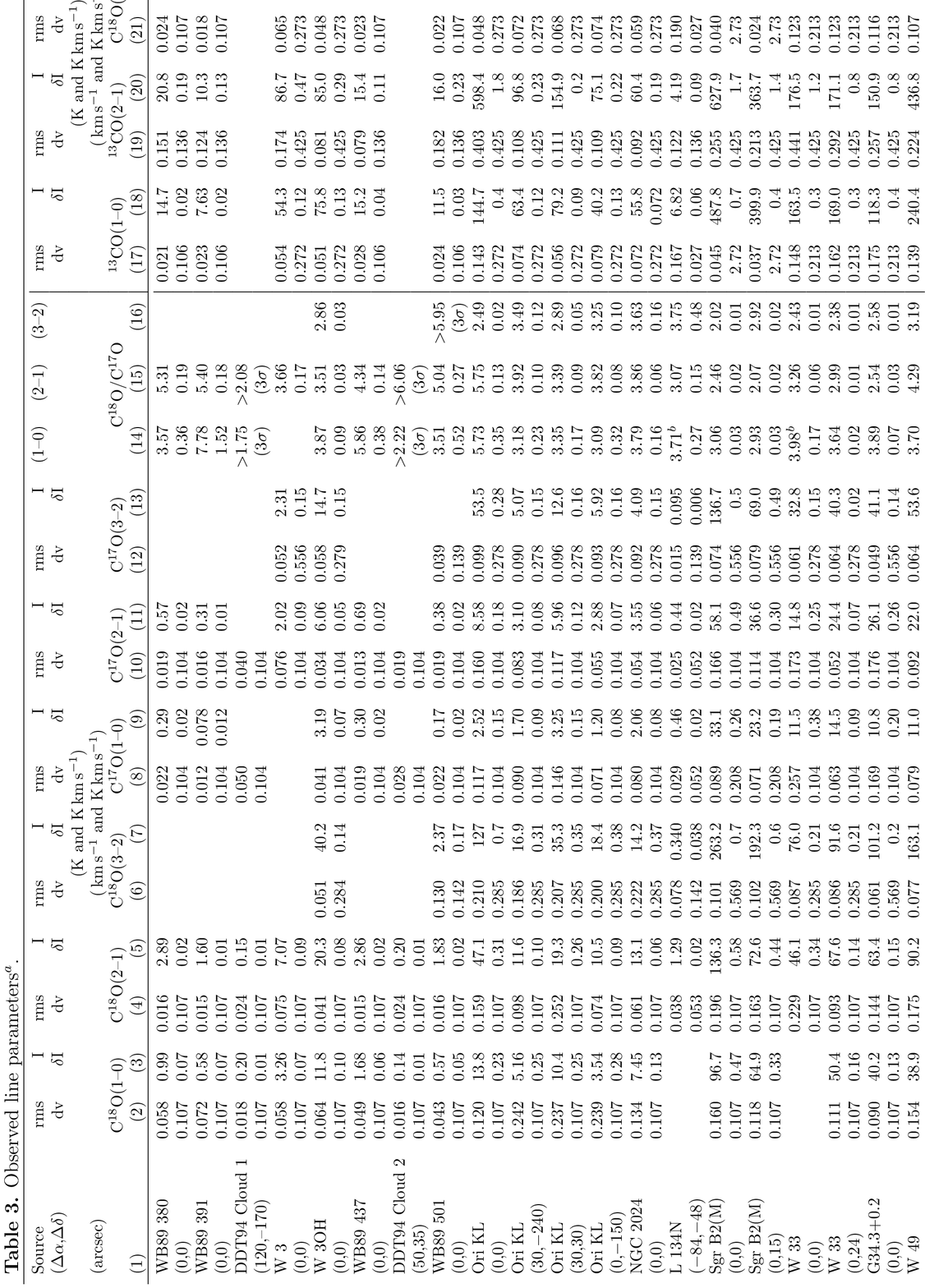}}}
\end{table*}

The resulting C$^{18}$O/C$^{17}$O abundance ratios as a function of distance 
from the galactic centre, $R_{\rm{GC}}$ are shown in Fig.~\ref{ratios_r_1}. 
Error bars accounting for the uncertainty in the line areas (obtained from 
baseline fits and given in Table~\ref{results}), are not plotted to avoid 
confusion. For comparison we also show the results from Penzias 
(\cite{penzias}) for the $J$=1--0 transition. For the $J$=2--1 lines, we plot 
results from Hofner 
et al.  (\cite{hofner}), who used the 30-m telescope to observe C$^{18}$O and 
C$^{17}$O towards H{\sc ii} regions of the inner Galaxy as well as towards 
W3(OH). The good agreement between our results and those of Penzias 
(\cite{penzias}) and Hofner et al. (\cite{hofner}) indicates that the 
calibration of our data is correct.

The distance to a given source can lead to a number of observational biases 
that are difficult to quantify and that do not facilitate a comparison between 
different parts of the Galaxy. With increasing distance, the linear beam size 
increases, leading potentially to the inclusion of a significant amount of 
relatively diffuse optically thin gas. On the other hand, a tendency to 
include brighter and more massive sources at larger distances may involve 
sources with systematically higher C$^{18}$O opacities that may lead to 
underestimates of the C$^{18}$O/C$^{17}$O abundance ratios. To 
search for a distance-related bias, Fig.~\ref{ratios_r_1r} therefore
shows the C$^{18}$O/C$^{17}$O ratios as a function of solar distance $d$. 
In Fig.~\ref{ratios_r_2} we show the same as in Fig.~\ref{ratios_r_1}, but 
for the $^{13}$CO to C$^{18}$O ratios obtained from the $J$=1--0 and 2--1 
lines. In addition, Figs.~\ref{ratios_r_2}c and d show correlations between 
the $^{13}$CO/C$^{18}$O and C$^{18}$O/C$^{17}$O ratios.

\section{Discussion}

\subsection{Overall trends}

Not accounting for optical depth effects and omitting Sgr~B2 and the far 
outer Galaxy clouds, our average C$^{18}$O/C$^{17}$O abundance ratios 
are 3.76$\pm$0.16 ($J$=1--0), 3.65$\pm$0.16 ($J$=2--1), and 3.08$\pm$0.12 
($J$=3--2), which can be compared with the Penzias (\cite{penzias}) $J$=1--0 
galactic disc value of 3.65$\pm$0.15. The $J$=1--0 results agree within 
the errors. If we weight our data with their uncertainties, the average 
ratios become slightly smaller, i.e. 3.68$\pm$0.11, 3.38$\pm$0.13, and 
2.74$\pm$0.10, respectively. We note, however, that smaller values 
tend to have smaller errors. Therefore, weighted averages may yield too 
small values and unweighted averages should be preferred. For the two 
SgrB2 positions average ratios are 2.99$\pm$0.06 (1--0), 2.26$\pm$0.17 
(2--1), and 2.47$\pm$0.38 (3--2) (weighted 2.99$\pm$0.05, 2.27$\pm$0.14, 
and 2.32$\pm$0.30). The average ratios of the four far outer Galaxy clouds 
are 5.18$\pm$0.96 (1--0) and 5.02$\pm$0.22 (2--1) (weighted: 4.79$\pm$0.77 
and 4.96$\pm$0.23). 

\begin{figure}
  \resizebox{\hsize}{!}{\rotatebox{0}{\includegraphics{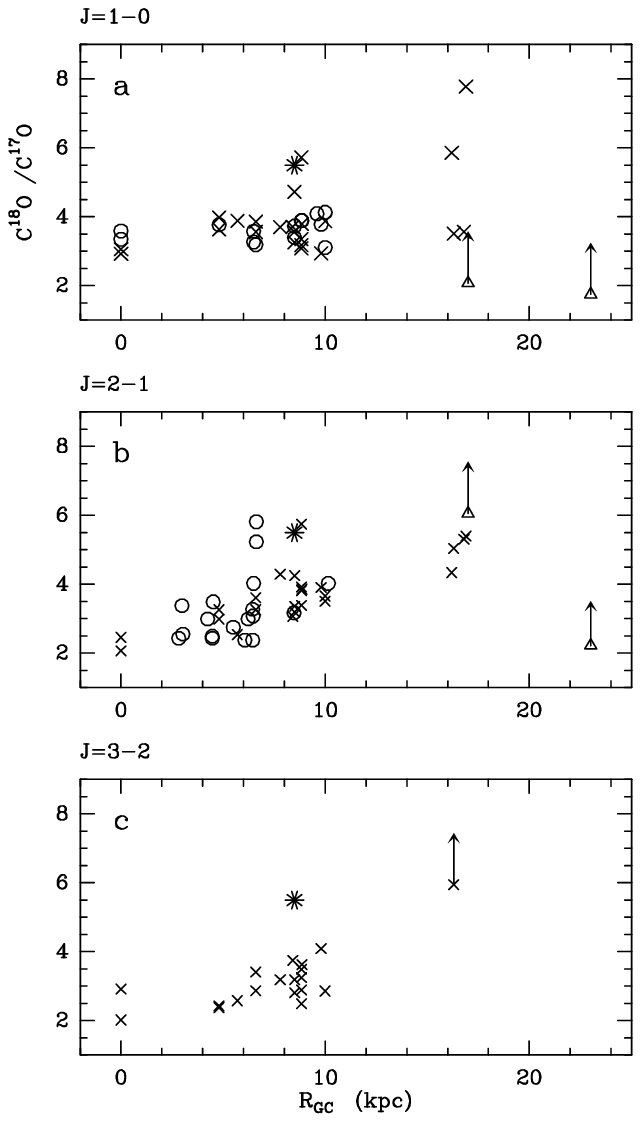}}}
 \caption{C$^{18}$O/C$^{17}$O abundance ratios, corrected for the difference
between the C$^{18}$O and C$^{17}$O frequencies but not accounting for 
optical depth effects, in the {\bf {a)}} $J$ = 1--0, {\bf {b)}} 2--1, 
and {\bf {c)}} 3--2 transition as a function of distance from the galactic 
centre ($R_{\rm{GC}}$). Crosses show our new results. Open triangles 
indicate 3$\sigma$ lower limits for DDT94 Cloud~1 and 2 that motivated 
this study. Open circles represent $J$=1--0 line results from Penzias 
(\cite{penzias}) in {\bf a)} and $J$=2--1 line results from Hofner et al.
(\cite{hofner}) in {\bf b)}. Error bars (see Table~\ref{results}) have 
been omitted to avoid confusion. The solar system value of 5.5 is marked
by an asterisk. }
 \label{ratios_r_1}
\end{figure}

\begin{figure}
  \resizebox{\hsize}{!}{\rotatebox{-90}{\includegraphics{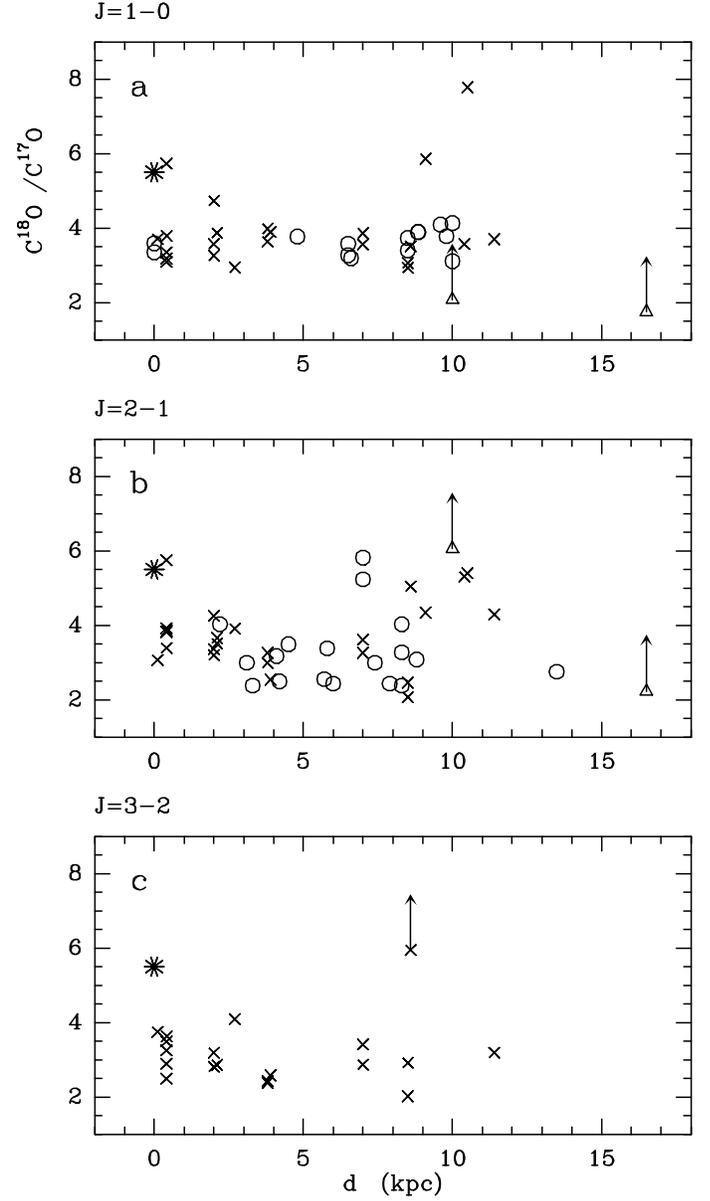}}}
 \caption{The same as Fig.~\ref{ratios_r_1}, but with distance from the
Sun plotted instead of $R_{\rm GC}$.}
 \label{ratios_r_1r}
\end{figure}

Assuming that the C$^{18}$O and C$^{17}$O rotational lines are optically thin, 
the most important results that can be derived from Figs.~\ref{ratios_r_1} 
and \ref{ratios_r_1r} are: (1) {\it Our C$^{18}$O/C$^{17}$O abundance ratios 
of the inner galactic disc are consistent with those of Penzias 
(\cite{penzias}), while the ratios from the galactic centre region are 
smaller}. 
(2) {\it There probably is a gradient with increasing $^{18}$O/$^{17}$O ratios 
as a function of $R_{\rm GC}$ out to galactocentric distances of 16\,kpc}. 
(3) {\it The solar system value is significantly larger than the ratios 
from the inner Galaxy, but not much higher than those from the far outer 
Galaxy}. (4) {\it An observational bias due to distance related effects 
is not apparent}.
 
\begin{figure}
  \resizebox{\hsize}{!}{\includegraphics{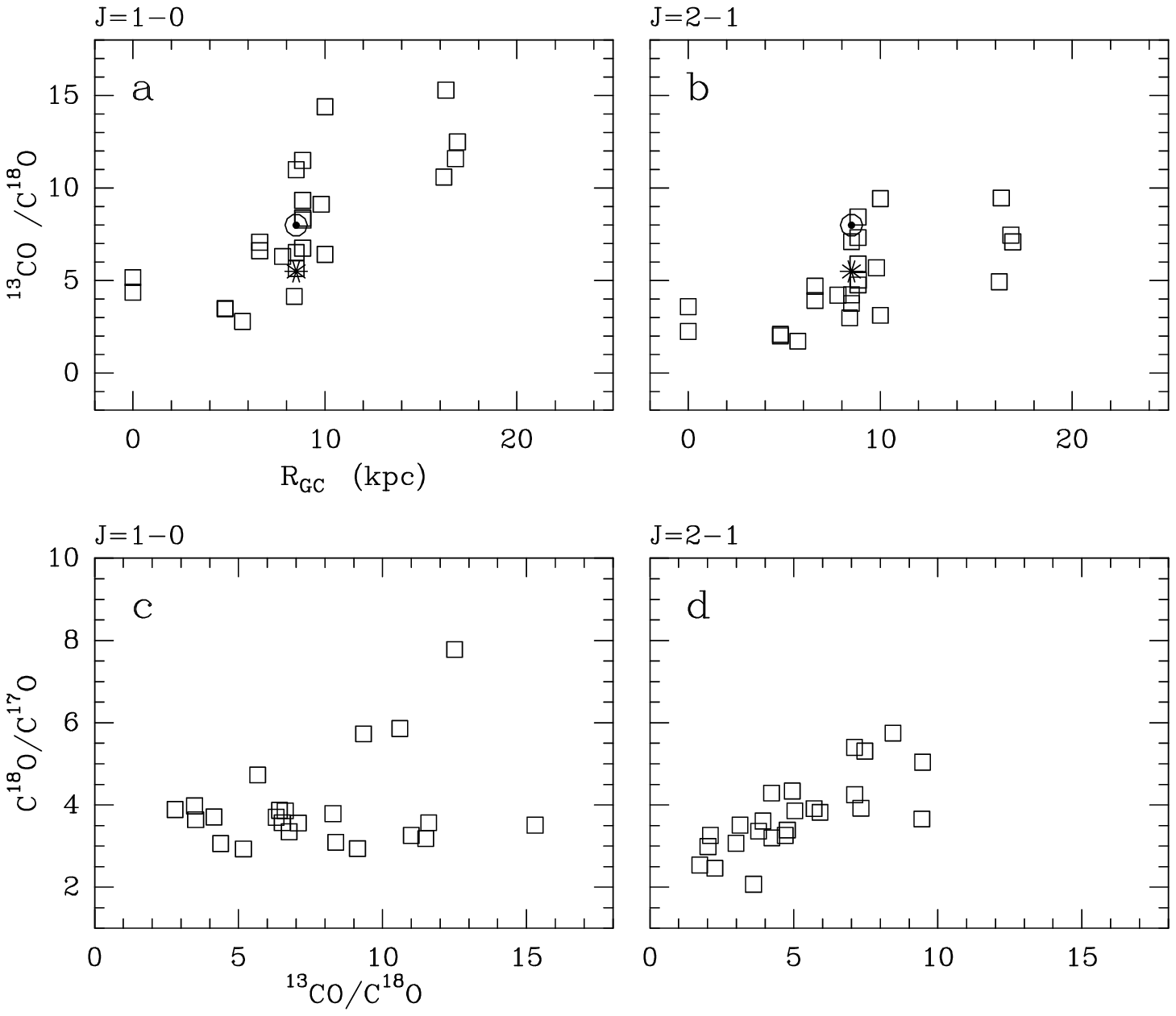}}
 \caption{$^{13}$CO/C$^{18}$O integrated intensity ratios in the 
{\bf {a)}} $J$ = 1--0 and {\bf {b)}} 2--1 transition as a function 
of distance from the galactic centre ($R_{\rm{GC}}$). Open squares 
indicate our new results. The values for the solar system of 5.5 (circle 
with central dot) and for the solar neighbourhood of 8.0 (asterisk) 
are also indicated. {\bf{c)}} The C$^{18}$O/C$^{17}$O 
integrated intensity ratio as a function of $^{13}$CO/C$^{18}$O for 
the $J$=1--0 and\ {\bf d)} $J$=2--1 transition. }
 \label{ratios_r_2}
\end{figure}

Fig.~\ref{ratios_r_1}a shows an increasing scatter in the $J$=1--0 ratios 
as a function of $R_{\rm{GC}}$. While in the inner Galaxy ratios are $\la$4,
there are a few higher ratios observed in clouds near the solar circle. 
Two (50\%) of the four far outer Galaxy (FOG) clouds are also characterised 
by a high ratio. The $J$=2--1 ratios show a more systematic increase with 
$R_{\rm{GC}}$ from about two near the galactic centre to about five at 
$R_{\rm{GC}}$=16~kpc. The $J$=3--2 ratios may follow this trend, but the 
ratios are generally smaller. The different behaviour of the $J$=1$-$0, 
2$-$1, and 3$-$2 transitions may be caused by differences in excitation and 
line opacities. Partial saturation of the stronger C$^{18}$O lines would lead 
to underestimated C$^{18}$O/C$^{17}$O abundance ratios. Since LVG calculations 
indicate $\tau_{(2-1)}$ $>$ $\tau_{(1-0)}$ for 
a wide range of parameters, saturation should be more pronounced in the 
2--1 transitions, which may explain some of the very low ratios obtained 
with the 2--1 lines for sources of the inner Galaxy. Clouds heated by 
young massive stars (this holds for the overwhelming majority of our sources, 
see Table~\ref{sources}) should be warm enough to yield $\tau_{3-2}$
$>$ $\tau_{2-1}$ (the $J$=3 state is located $\sim$30\,K above the $J$=0 
level) which readily explains the relatively small $J$=3--2 
C$^{18}$O/C$^{17}$O ratios, which are plotted in Fig.\,\ref{ratios_r_1}c.

Gradients may also be present in the $^{13}$CO/C$^{18}$O $J$=1--0 and 2--1
ratios (Figs.\,\ref{ratios_r_2}a and b). 
If real, this can be explained by a lack of diffuse 
gas in the high pressure environment of the galactic centre region, while
sources further out may show large amounts of fractionated diffuse gas with
enhanced $^{13}$CO abundances (e.g., Watson et al. \cite{watson}; Bally \&
Langer \cite{bally}). 

To directly determine optical depths, we fitted the relative 
strengths of the two C$^{17}$O\,(1--0) groups of hyperfine components 
for sources with sufficiently small linewidths. In most cases, i.e. 
for WB89~391, WB89~437, WB89~501, Ori-KL\,(30,--240), and L134N, the optical 
depth of the C$^{17}$O(1--0) line is $\la$0.1. Higher values for the 
C$^{17}$O\,(1--0) optical depth were found towards WB89~380 (0.78$\pm$0.72), 
NGC~2024 (0.82$\pm$0.38; this source shows more than one velocity component, 
but a two-component fit did not provide a reliable value), Ori-KL\,(30,30) 
(0.76$\pm$0.34), and Ori-KL\.(0,$-$150) (0.27$\pm$0.53). In all of these 
cases, uncertainties are high. A one velocity component fit to the 
C$^{17}$O(2--1) transition could only be obtained towards L134N: 
$\tau$=0.53$\pm$0.11. Assuming $\tau[$C$^{18}$O(1--0)] $\sim$ 
4$\times$$\tau[$C$^{17}$O(1--0)], these results suggest that 
$\tau($C$^{18}$O(1--0)) is commonly {\it but not always} smaller 
than unity and that $\tau$[C$^{18}$O(2--1)] $>$ $\tau$[C$^{18}$O(1--0)]. 
Figs.\,\ref{ratios_r_2}c supports this finding, indicating no saturation
effects in the $J$=1--0 lines. C$^{18}$O/C$^{17}$O appears to be independent 
of $^{13}$CO/C$^{18}$O. The opposite appears, however, to hold for the 2--1 
lines. Here small $^{13}$CO/C$^{18}$O ratios are accompanied by small
C$^{18}$O/C$^{17}$O values, suggesting some degree of $^{13}$CO and 
C$^{18}$O $J$=2--1 line saturation. 

Towards all sources we observed one or a few positions only. A map of the 
Orion-KL cloud was, however, made by White \& Sandell (\cite{white}) in
the C$^{18}$O and C$^{17}$O (2--1) transitions. Although their primary goal 
was not the determination of $^{18}$O/$^{17}$O abundance ratios, their 
Fig.\,8 might indicate a surprising trend towards larger C$^{18}$O/C$^{17}$O
ratios at larger extinctions or C$^{18}$O column densities that deserves 
further 
study. Our value for Ori-KL\,(0,0) (both for the $J$=1--0 and 2--1 lines) is 
consistent with the ratios found by White \& Sandell (\cite{white}) at high 
$A_{\rm{v}}$.

\subsection{LVG calculations}

\subsubsection{Models with C$^{18}$O/H$_2$ = 1.7$\times$10$^{-7}$}

For all sources the combined data sets (excluding $^{13}$CO, because some of 
it probably originates from a more diffuse lower density environment) have 
been analysed with an LVG code, in the same way as done by Wouterloot 
et al. (\cite{wbh05}) for the $\rho$~Oph cloud. We used rates by Flower 
(\cite{flower}) for collisions with H$_2$ and assumed an ortho/para H$_2$
abundance ratio of three (our results are not sensitively dependent on
this value). For L134N, a cold dark cloud, the same parameter space was 
used as for $\rho$~Oph in Wouterloot et al. (\cite{wbh05}) (5~K $\leq$ 
$T_{\rm kin}$ $\leq$ 35~K, 10$^3$~cm$^{-3}$ $\leq$ $n$(H$_2$) $\leq$ 
10$^6$~cm$^{-3}$, and 2.0 $\leq$ C$^{18}$O/C$^{17}$O $\leq$ 6.0). For 
the other sources $T_{\rm kin}$ ranged from 5~K to 100~K or even 130~K 
(for DR21cal) using the same range of densities and C$^{18}$O/C$^{17}$O 
ratios as for L134N. In all cases the adopted velocity gradient was 
5.0\,kms$^{-1}$pc$^{-1}$ that roughly reflects cloud size and measured 
linewidths. 

\begin{figure}
 \resizebox{\hsize}{!}{\rotatebox{-90}{\includegraphics{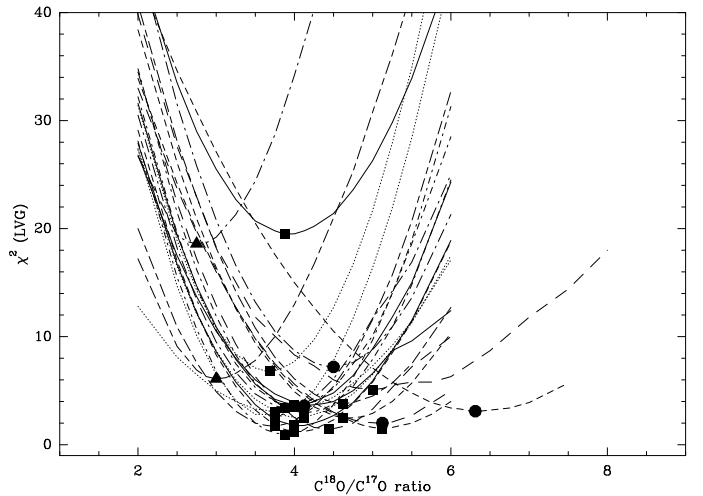}}}
 \caption{Results of LVG simulations modeling our sample of sources: $\chi^2$ 
as a function of assumed C$^{18}$O/C$^{17}$O ratio. The minimum $\chi^2$
values are indicated by filled triangles (Sgr~B2), squares 
(3kpc$<$$R_{\rm GC}$$<$10kpc), 
and circles ($R_{\rm GC}$$>$16kpc). The calculations were made with steps of 
0.125 (near the minimum $\chi^2$) to 0.50 in the C$^{18}$O/C$^{17}$O 
ratio.  The C$^{18}$O/H$_2$ abundance ratio was assumed to be 1.7~10$^{-7}$ 
(Frerking et al.  \cite{frerking}). The results are similar to those also 
accounting for a galactic C$^{18}$O/H$_2$ abundance gradient.}
\label{lvgfrerking}
\end{figure}

The models were compared with the observations by calculating $\chi^2$ 
values from the C$^{18}$O peak temperatures and the C$^{18}$O/C$^{17}$O 
ratios, using for the latter integrated line temperatures. For this procedure, 
the observed peak $T_{\rm{A}}^*$ temperatures of C$^{18}$O were converted 
to $T_{\rm{mb}}$ temperatures except for L134N, where the emission is much 
less peaked at the observed position than for the other sources (see e.g., 
Pratap et al. \cite{pratap}). The $J$=2--1 and 3--2 peak temperatures used 
are those of spectra that were convolved to the $J$=1--0 resolution using 
results from the C$^{18}$O\,(2--1) HERA maps. For the uncertainties we 
assumed a calibration error of $\pm$10\% in the line temperatures and of 
$\pm$14\% in the line ratios (see Sect.\,2.3). In Fig.~\ref{lvgfrerking}, 
$\chi^2$ is shown as a function of the adopted C$^{18}$O/C$^{17}$O ratio. 
This ratio was varied in steps of 0.125 (near the minimum $\chi^2$) to
0.5. For 
these models we assumed the Frerking et al. (\cite{frerking}) C$^{18}$O/H$_2$ 
abundance ratio of 1.7~10$^{-7}$. All sources show a well-defined minimum. 
The parameters of the model with the lowest $\chi^2$ are listed in 
Table~\ref{lvgresults}, Col. 4.

\addtocounter{table}{1}

%
%
\begin{table*}
\caption[]{Results of LVG model calculations$^{a}$.
\label{lvgresults}}
\begin{flushleft}
\begin{tabular}{clrccccccccr}
\hline\noalign{\smallskip}
Nr & Source & Offset & C$^{18}$O/C$^{17}$O & $T_{\rm kin}$ & log\,$n$(H$_2$) & $\chi^2$ & 
C$^{18}$O/C$^{17}$O & $T_{\rm kin}$ & log\,$n$(H$_2$) & $\chi^2$ & $R_{\rm GC}$ \\
   &    &  (arcsec)    &                    &   (K)  & (cm$^{-3}$) &                     &   
   & (K)  & cm$^{-3}$ & & kpc \\
 & & & \multicolumn{4}{c}{C$^{18}$O/H$_2$ Frerking}&
\multicolumn{4}{c}{C$^{18}$O/H$_2$ Frerking+gradient} \\
\hline\noalign{\smallskip}
1& WB89 380 & 0,0     &  4.13 & 77.5 &  4.0  & 3.6 & 4.13 & 45.0 & 4.8 & 3.4 & 16.8 \\
2& WB89 391 & 0,0     &  6.31 & 70.0 &  4.1  & 3.1 & 6.38 & 45.0 & 4.9 & 2.6 & 16.9 \\
3& W 3      & 0,0     &  3.75 & 25.0 &  4.3  & 3.0 & 3.88 & 25.0 & 4.5 & 6.2 & 10.0 \\
4& W 3OH    & 0,0     &  3.88  & 30.0 &  4.8  & 0.9 & 3.88 & 27.5 & 4.9 & 1.6 & 10.0 \\
5& WB89 437 & 0,0     &  5.13  & 30.0 &  4.0  & 2.0 & 5.13 & 20.0 & 4.7 & 2.4 & 16.2 \\
6& WB89 501 & 0,0        &  4.5  & 60.0 &  3.9  & 7.2 & 4.5 & 20.0 & 4.7 & 16.4 & 16.3 \\
7& Ori KL   & 0,0     &  3.88  & 52.5 &  4.9  & 19.5& 3.88 &  62.5 & 5.0 & 19.1&  8.84 \\
8& Ori KL   & 30,30   &  4.0  & 45.0 &  5.3  & 1.2 & 4.13 & 42.5 & 5.3 & 2.1 &  8.84 \\
9& Ori KL   & 0,--150 &  3.75 & 45.0 &  4.9  & 2.4 & 3.75 & 52.5 & 5.0 & 2.4 &  8.84 \\
10&Ori KL   & 30,--240&  4.0  & 40.0 &  5.0  & 3.5 & 4.0 & 37.5 & 5.0 & 4.8 &  8.84 \\
11&NGC 2024 & 0,0        &  4.63 & 20.0 &  4.8  & 3.8 & 4.5 & 20.0 & 4.8 & 5.7 &  8.86 \\
12&L 134N &   0,0        &  5.0 &  7.0 &  4.7  & 5.1 & 5.0 & 7.0 & 4.6 & 5.6 &  8.41 \\
13 &Sgr B2M  & 0,0   &  3.0  & 37.5 &  5.1  & 6.1 & 3.0 & 42.5 & 4.3 & 5.3 & 0.1 \\
14 &Sgr B2M  & 0,15  &  2.75 & 35.0 &  4.8  & 18.6& 2.75 & 57.5 & 3.9 & 21.5 &  0.1 \\
15&W33      & 0,0     &  4.0  & 30.0 &  5.1  & 1.8 & 4.25 & 30.0 & 4.8 & 1.1 &  4.8 \\
16&W33      & 0,24    &  5.13 & 30.0 &  5.5  & 1.5 & 4.75 & 32.5 & 5.0 & 1.6 &  4.8 \\
17&G34.3+0.2& 0,0        & 4.13  & 35.0 &  5.3  & 2.5 & 4.13 & 35.0 & 5.0 & 2.2 &  5.7 \\
18&W49N     & 0,0        &  4.13 & 50.0 &  5.0  & 3.0 & 4.25 & 45.0 & 4.9 & 3.2 &  7.8 \\
19&W51d-a$^b$& 0,0       &  4.44  & 12.5 &  4.7  & 1.5 & 4.5 & 12.5 & 4.4 & 1.3 &  6.6 \\
20&W51d-b$^b$& 0,0       &  3.88  & 25.0 &  4.8  & 3.4 & 4.0 & 25.0 & 4.7 & 3.6 & 6.6 \\
21&DR 21    & 0,0        &  3.75 & 107.5 & 4.7  & 1.7 & 3.75 & 107.5 & 4.7 & 1.7 &  8.5 \\
22&DR 21cal & 0,0        &  4.0  & 30.0 &  5.0  & 3.7 & 4.0 & 30.0 & 5.0 & 3.7  &  8.5 \\
23&DR 21OH  & 0,0        & 4.63 & 22.5 &  4.9  & 2.5 & 4.63 & 22.5 & 4.9 & 2.5 &  8.5 \\
24&NGC 7538 & 0,0        & 3.69 & 40.0 &  4.8  & 6.8 & 3.63 & 40.0 & 4.9 & 6.7 &  9.8 \\
\noalign{\smallskip}
\hline
\end{tabular}
\end{flushleft}
$^{a}$: Cols. 2 and 3: source 
and offset; Cols.\,4 to 7: derived C$^{18}$O/C$^{17}$O ratio, kinetic 
temperature, H$_2$ density and minimum $\chi^2$. \\
Cols.\,8 to 11: the same parameters as Cols.\,4 to 7, but for the case of a 
galactic C$^{18}$O/H$_2$ abundance gradient (see Sects.\,4.1 and 4.2).\\
$^{b}$: W51d-a: 50\,km\,s$^{-1}$ component; W51d-b: 60\,km\,s$^{-1}$ component 
(see Fig.\,\ref{spectra}).
\end{table*}

The average C$^{18}$O/C$^{17}$O ratio is 4.18$\pm$0.15. Excluding the 
two  positions towards Sgr~B2, which have low ratios (2.88$\pm$0.11), and 
the objects at $R_{\rm GC}$$>$16kpc, which have high values (5.02$\pm$0.45), 
the average ratio becomes 4.15$\pm$0.10. This is close to the values 
(4.11$\pm$0.14 and 4.13$\pm$0.13) derived for $\rho$~Oph and NGC\,6334 FIR\,II 
by Wouterloot et al.  (\cite{wbh05}) and Zhang et al. (\cite{zhang}), 
respectively.

\subsubsection{Introducing a C$^{18}$O abundance gradient}

From optical and FIR observations it has been shown that many metals show
a considerable abundance gradient. Abundances decrease from the galactic 
centre to larger $R_{\rm GC}$ (e.g., Rudolph et al. \cite{rudolph}). 
Gradients amount to $-$0.078\,dex/kpc for $^{14}$N/H, $-$0.051\,dex/kpc 
for $^{16}$O/H, and $-$0.044\,dex/kpc for $^{32}$S/H (these are average 
values from optical and FIR measurements which yield slightly different 
results). For $^{12}$C/H, there are only few data. Rolleston et al. 
(\cite{rolleston}) obtained a slope of $-$0.070 dex/kpc from 80 stars in 
19 open clusters (6$\le$$R_{\rm GC}$$\le$18kpc). Esteban et al. (\cite{esteban})
observed recombination lines towards 8 H{\sc ii} regions (6.3 $\le$ 
$R_{\rm GC}$ $\le$ 10.4kpc) and derived a slope of $-$0.103 dex/kpc. 
Model calculations by Matteucci \& Fran\c{c}ois (\cite{matteucci}) 
suggest a gradient of $-$0.066 dex/kpc for $^{12}$C/H. In all cases, 
where measurements extend to large galactocentric radii, there is no sign 
for a change in the radial gradient. 

In addition to the gradients in the $^{12}$C and $^{16}$O abundances, 
there is also a gradient in the $^{16}$O/$^{18}$O ratio: Wilson \& Rood 
(\cite{wilson}) give a ratio of $\sim$250 for the galactic centre region 
and a fitted ratio of $^{16}$O/$^{18}$O $\sim$ (59$\pm$12)$\times$$R_{\rm GC}$
+ (37$\pm$83) for larger $R_{\rm GC}$ values (2.8 $\leq$ $R_{\rm GC}$ $\leq$ 
9~kpc).

Adopting the above mentioned $^{16}$O/H abundance gradient of 
$-$0.051\,dex/kpc, the oxygen abundance becomes a factor 
of 2.7 higher near the galactic centre and a factor of 2.4 
lower at $R_{\rm GC}$ = 16\,kpc than in the solar neighbourhood. Similarly, 
the $^{16}$O/$^{18}$O ratio is then factor of 2.1 lower 
near the centre (using a minimum $^{16}$O/$^{18}$O ratio of 250) and a 
factor of 1.8 higher at $R_{\rm GC}$ = 16\,kpc than the fitted value near 
the Sun, {537 (Wilson \& Rood \cite{wilson}) which can be compared with
the solar system value of 490 (Anders \& Grevesse \cite{anders}). 

Applying both gradients to estimate $^{18}$O/H, the ratios become a 
factor of 5.7 higher at the centre and a factor of 4.4 lower at 
$R_{\rm GC}$ = 16\,kpc than in the solar neighbourhood. Taking for $^{12}$C
the mean of the gradients determined by Rolleston et al. (\cite{rolleston})
and Esteban et al. (\cite{esteban}), \hbox{--0.0865}\,dex/kpc, we obtain 
$^{12}$C/H abundances that vary between 5.4 higher and 4.5 times lower within
the same range of galactocentric radii. Since the gradients are uncertain and 
the true abundances also depend on chemical reactions in the clouds, the 
$^{18}$O/H 
and $^{12}$C/H gradients are identical within the error limits. In our LVG 
calculations we have therefore used the $^{18}$O/H gradient for all sources 
to obtain a second estimate of $^{18}$O/$^{17}$O ratios. For large 
$R_{\rm GC}$, the extrapolated lower abundances are consistent with chemical 
model calculations (Ruffle et al. \cite{ruffle}) simulating multi-line
measurements of DDT94~Cloud 2.

\begin{figure}
  \resizebox{\hsize}{!}{\includegraphics{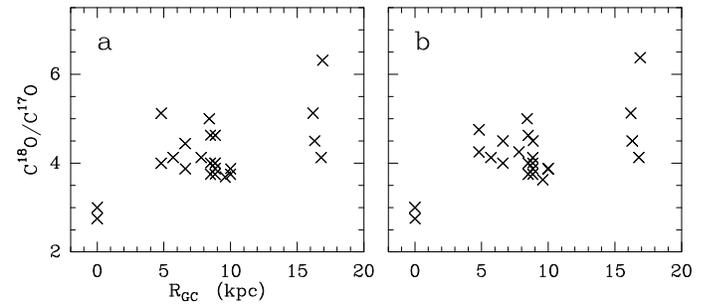}}
 \caption{Results of LVG calculations: The C$^{18}$O/C$^{17}$O ratio as a 
function of $R_{\rm GC}$. {\bf{a)}} The C$^{18}$O/H$_2$ ratio was assumed to 
be 1.7~10$^{-7}$ (Frerking et al. \cite{frerking}), or {\bf{b)}} corrected 
for the galactic abundance gradients (see Sect.\,4.2.2).}
 \label{lvggradient}
\end{figure}

The results are given in Table~\ref{lvgresults} (Cols.\,7 to 10) and 
Fig.~\ref{lvggradient}. There is little change in the C$^{18}$O/C$^{17}$O 
ratios compared to those without considering the abundance gradient 
(Col.\,4 of Table~\ref{lvgresults}). The average C$^{18}$O/C$^{17}$O ratio 
becomes 4.20$\pm$0.14 instead of 4.18$\pm$0.15. Excluding 
the two positions in Sgr~B2, which have low ratios (2.88$\pm$0.11), and the 
objects at $R_{\rm GC}$$>$16kpc, which have higher values (5.03$\pm$0.46), 
the average ratio becomes 4.16$\pm$0.09. This is again close to the value 
derived by Wouterloot et al. (\cite{wbh05}) for $\rho$~Oph (see also Zhang et 
al. \cite{zhang} for NGC6334 FIR\,II). There are, however, differences in 
the derived kinetic temperatures and densities. In the first set of models 
(with fixed C$^{18}$O/H$_2$ abundance ratios) the derived densities are 
significantly smaller at $R_{\rm GC}$$>$16kpc than for 
3$<$$R_{\rm GC}$$<$10kpc. 
This difference is greatly reduced in the second set of models accounting 
for C$^{18}$O/H$_2$ gradients (Fig.~\ref{ntcomp}). If we assume that the H$_2$ 
densities in star forming clumps such as those in our selected sample are 
similar at all locations in the Galaxy, this indicates that the models with 
the abundance gradient are more realistic.

\begin{figure}
  \resizebox{\hsize}{!}{\rotatebox{-90}{\includegraphics{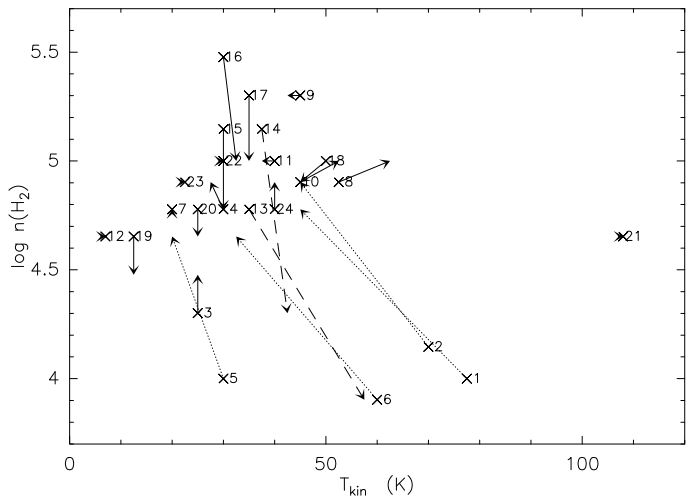}}}
 \caption{Results of LVG calculations: The crosses indicate the best fit
$T_{{\rm kin}}$ and $n$(H$_2$) using the Frerking et al. (\cite{frerking}) 
abundance. The arrows point to the results for the model using an abundance 
gradient. The arrows for the clouds at R$>$16kpc are dotted, those for 
Sgr~B2 are dashed. Numbers indicate the sequential line-of-sight number, 
indicated in Table~\ref{lvgresults}.
}
 \label{ntcomp}
\end{figure}

\subsubsection{Implications}

The two sets of models, either ignoring or accounting for radial
C$^{18}$O/H$_2$ abundance gradients across the Galaxy, suggest the existence 
of a C$^{18}$O/C$^{17}$O gradient from values slightly smaller than three 
near the galactic centre to values around five at large galactocentric radii 
(Fig.~\ref{lvggradient}). The $^{18}$O/$^{17}$O ratio of Sgr~B2 deduced 
from HCO$^+$, 3.1$\pm$0.6 (Gu{\'e}lin et al. \cite{guelin}), is consistent 
with this result. Our results do suggest that the C$^{18}$O/C$^{17}$O ratio 
at 4 to 11\,kpc from the galactic centre is larger than the value derived 
by Penzias (\cite{penzias}; 3.65$\pm$0.15). The reason is that the $J$=1--0 
results from Penzias were not corrected for radiative transfer effects.  

$^{18}$O is synthesised by helium burning on $^{14}$N in massive stars. 
$^{14}$N is a mostly secondary nucleus so that its abundance is highly 
metallicity-dependent, with low N/C and N/O abundance ratios at low 
metallicities (e.g., Wheeler et al. \cite{wheeler}). $^{18}$O may follow 
this trend. $^{17}$O 
is mainly a product of CNO-hydrogen burning and is also released by stars of 
intermediate mass (for yields, see, e.g., Prantzos et al. \cite{prantzos}).
 
To qualitatively reproduce the $^{18}$O/$^{17}$O ratios observed in 
interstellar clouds and stellar objects, we should note that in individual 
dust grains from massive stars $^{18}$O overabundances of up to two orders of 
magnitude  can be reached (e.g., Amari et al. \cite{amari}). Low ratios, 
sometimes below unity, are obtained towards late-type stars of intermediate 
mass (c.f., Sect.\,1). The low values indicate that $^{18}$O is destroyed in 
such stars (see also Henkel \& Mauersberger \cite{henkelmau}; Henkel et al. 
\cite{henkelwil}). 

For the Large Magellanic Cloud (LMC), the C$^{18}$O/C$^{17}$O\,(2--1) ratio 
has been studied by Heikkil\"a et al. (\cite{heikkila}). These 
Clouds 
have even lower metallicities than the far outer Galaxy and are
characterised by C$^{18}$O/C$^{17}$O $\sim$ 1.6, which is significantly lower 
than all measured interstellar values in the Galaxy. The LMC, which 
undergoes a phase of enhanced massive star formation, is metal-poor so that 
its $^{18}$O abundance should be low. Very high C$^{18}$O/C$^{17}$O ratios 
are instead found in nuclear starbursts (e.g., Harrison et 
al.~\cite{harrison}; Wang et al. \cite{wang}). This may be a consequence of 
(1) high 
metallicities and (2) large amounts of ejecta from massive stars that must 
have enriched the interstellar medium. 

In the central region of our Galaxy, metallicities are also high. However, 
here CNO burning dominates resulting isotope ratios of C, N, and O. 
Apparently, a nuclear starburst with large numbers of massive stars has not 
contaminated the galactic 
centre region since a long time. Products from helium burning are therefore
underepresented in the C, N, and O isotope abundances, yielding a very 
low $^{12}$C/$^{13}$C ratio ($\sim$25), an extremely large $^{14}$N/$^{15}$N 
ratio (of order 1000) (e.g., Wilson \& Rood \cite{wilson}) and the low 
$^{18}$O/$^{17}$O ratio ($\la$ 3) proposed here. 

What remains, at first sight, puzzling is (1) that there exists a galactic 
disc $^{18}$O/$^{17}$O gradient and (2) that the high $^{18}$O/$^{17}$O ratios 
in the metal-poor outer Galaxy do not match the low values in the even more 
metal-poor LMC. Does this latter discrepancy imply that the metallicities of 
the outer Galaxy still permit efficient high-mass star production of 
$^{18}$O?  While the LMC is too metal-poor to make the process efficient?  
Or that, in spite of the presently visible large numbers of massive stars 
the LMC starburst is still too young to enrich the interstellar medium with 
$^{18}$O-rich ejecta? Both possibilities appear to be farfetched. The 
first, because the metallicities are not that different (e.g., Hunter et al. 
\cite{hunter}; Ruffle et al. \cite{ruffle}) to justify an extreme variation 
(by a factor of three) between measured $^{18}$O/$^{17}$O ratios, the second, 
because the starburst in the LMC started as early as 50$\times$10$^6$\,yr ago 
(e.g., Westerlund \cite{westerlund}). 

The galactic gradient and the difference between outer Galaxy and LMC
may instead be interpreted in terms of current models of galacto-chemical 
evolution. The most accepted mechanism to explain the existence of abundance 
gradients is the so-called ``biased infall'' (e.g., Chiappini \& Mateucci 
\cite{chiappini}) with the galactic disc being slowly formed from inside out.
In such a scenario, $^{18}$O abundances may be low in both the outer disc and 
LMC and the difference is mainly caused by $^{17}$O. In the LMC, where star 
formation is ongoing since at least 10$^{10}$\,yr (e.g., Hodge \cite{hodge}), 
there was sufficient time to form $^{17}$O predominantly from stars of 
intermediate mass. The outer Galaxy, however, may be too young to build up
a similar $^{17}$O abundance, with timescales (and abundances) rapidly 
decreasing the farther out the measured molecular cloud is located.

To summarise, we {\it may have found a fundamental difference between 
the metal-poor outer regions of the Galaxy and the metal-poor LMC.}
Generalizing 
this result, the difference implies that the abundances of the outer regions 
of spirals cannot be considered to be intermediate between those found near 
the solar circle and those determined in small metal-poor dwarf galaxies. As a 
cautionary note, however, we have to emphasise that {\it the $^{18}$O/$^{17}$O
ratios of the innermost and outermost galactic star-forming regions are still 
based on a very small number of targets}. Additional data to confirm or to 
reject the trend of rising $^{18}$O/$^{17}$O ratios with galactocentric 
radius would thus be highly desirable. 

\section{Conclusions}

To measure interstellar $^{18}$O/$^{17}$O ratios, we observed three 
C$^{18}$O and C$^{17}$O transitions towards 25 positions in 18 galactic 
sources located at galactocentric distances 0\,kpc $\leq$ $R_{\rm GC}$ 
$\leq$ 16.9\,kpc. These measurements are complemented by $^{13}$CO 
observations 
in the two ground rotational transitions. C$^{18}$O/C$^{17}$O line intensity 
ratios from the $J$=2--1 and 3--2 transitions are lower than those from the 
1--0 lines, indicating that optical depth effects have to be considered. 
This is also suggested by a comparison with the $^{13}$CO (1--0) and (2--1) 
spectra. 

Therefore, all C$^{18}$O and C$^{17}$O observations were combined by means 
of LVG calculations. This was done assuming (1)
that the C$^{18}$O/H$_2$ abundance ratio is constant throughout the Galaxy 
and (2) that there is a radial C$^{18}$O/H$_2$ abundance gradient. These 
assumptions do not affect significantly the final C$^{18}$O/C$^{17}$O values, 
which should provide a good approximation to the $^{18}$O/$^{17}$O isotope 
ratio. For the central region, $^{18}$O/$^{17}$O = 2.88$\pm$0.11. For 
$R_{\rm GC}$ = 4--11\,kpc, the ratio becomes 4.16$\pm$0.09. In the 
outer Galaxy (16\,kpc $\leq$ $R_{\rm GC}$ $\leq$ 17\,kpc), we find 
5.03$\pm$0.46.

The low ratio in the galactic centre region is consistent with a CNO-hydrogen 
burning dominated nucleosynthesis that is also characterizing the carbon and 
nitrogen isotope ratios. It supports the view that $^{18}$O is predominantly 
synthesised in high-mass stars, while $^{17}$O is predominantly a product of 
stars of lower mass. The ratio between 4 and 11\,kpc is consistent with 
recent results from a few ``local'' individual clouds. It is smaller than the 
solar
system ratio (5.5), suggesting that the Sun was enriched by material from 
massive stars during its formation. The $^{18}$O/$^{17}$O values measured 
for the outer Galaxy are more difficult to interpret. They are not low, as 
in the case of the metal-poor massive star forming Large Magellanic Cloud 
but appear to be higher than anywhere else in the interstellar medium of the 
Galaxy. This and the galactic disc gradient may be explained by the small 
age of the outer galactic disc and may imply that the metal-poor outer reaches 
of spiral galaxies provide quite different environments than similarly 
metal-poor 
dwarf galaxies. Nevertheless, both the Galactic centre and outer Galaxy 
$^{18}$O/$^{17}$O isotope ratios suffer from small number statistics so that 
more observations are needed to confirm or to reject the trend found in this 
study.

\Online
\appendix

\section{Spectra}

\begin{figure*}
 \resizebox{14.5cm}{!}{\includegraphics{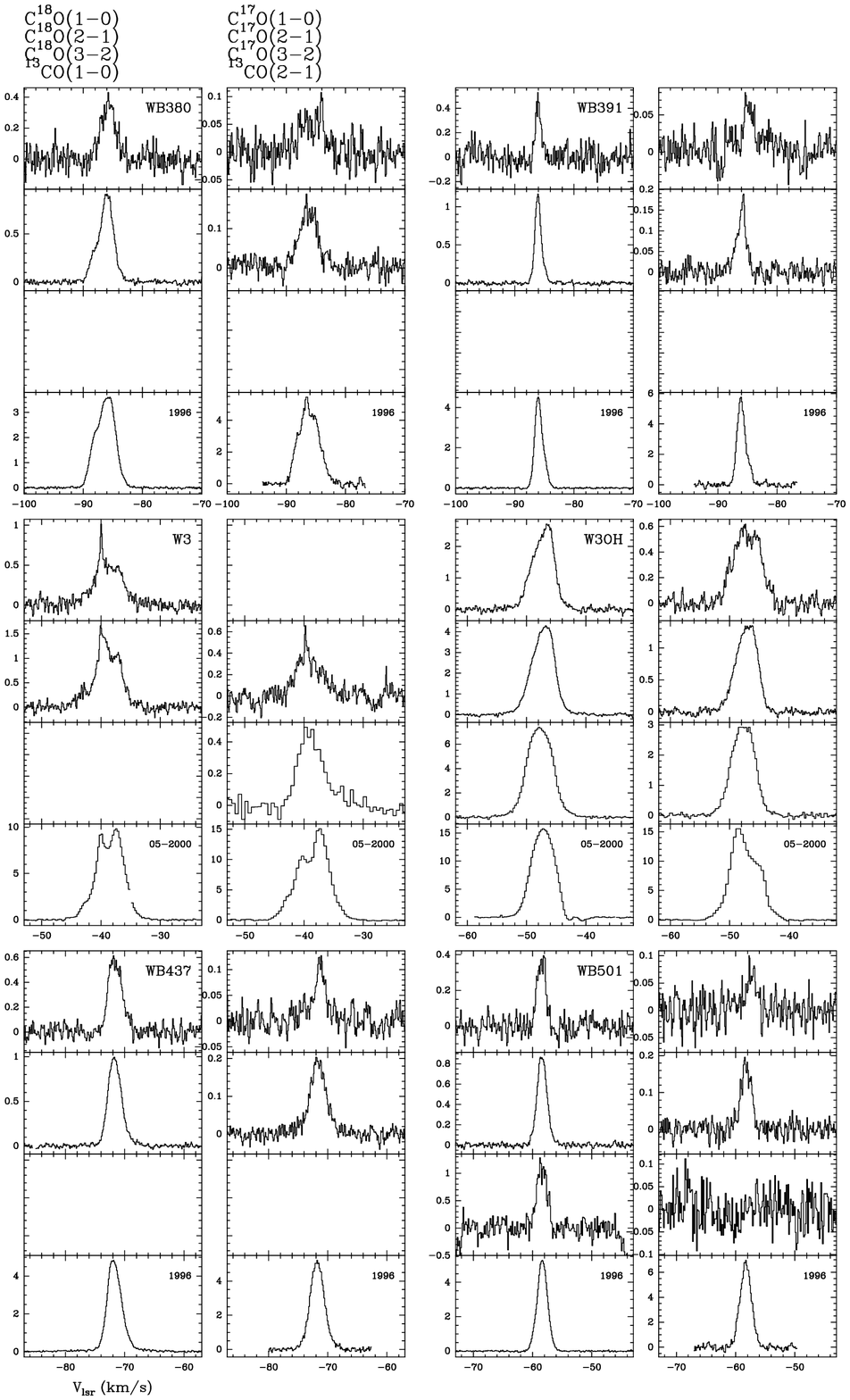}}
 \caption{C$^{17}$O, C$^{18}$O $J$ = 1--0, 2--1, and 3--2, and $^{13}$CO 
$J$ = 1--0 and 2--1 spectra. 
In most cases the velocity range is 30~km$\,$s$^{-1}$, except for a 
few sources with very broad lines.}
 \label{spectra}
\end{figure*}

\addtocounter{figure}{-1}

\begin{figure*}
 \resizebox{14.5cm}{!}{\includegraphics{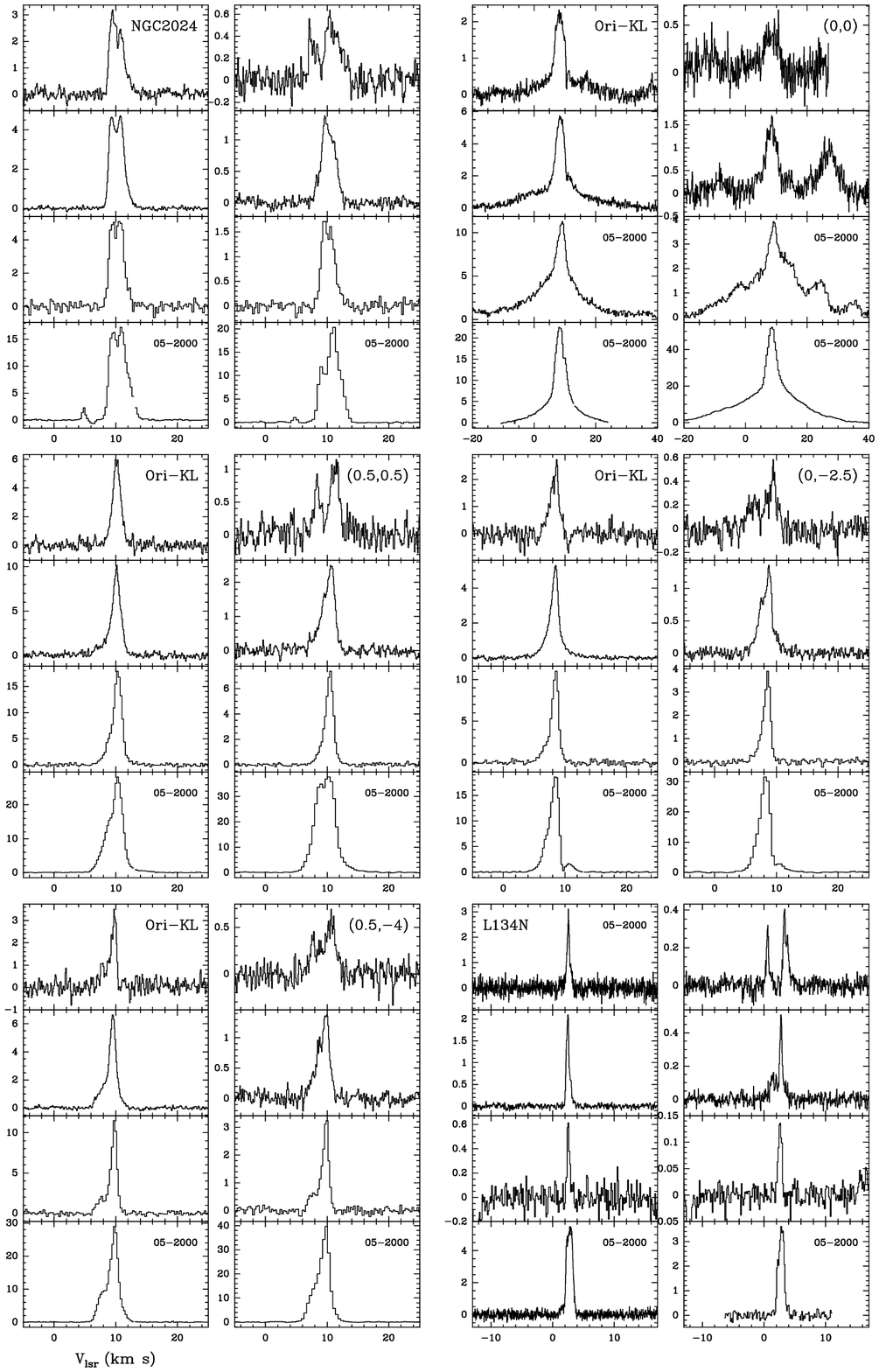}}
 \caption{Continued.}
\end{figure*}

\addtocounter{figure}{-1}

\begin{figure*}
 \resizebox{14.5cm}{!}{\includegraphics{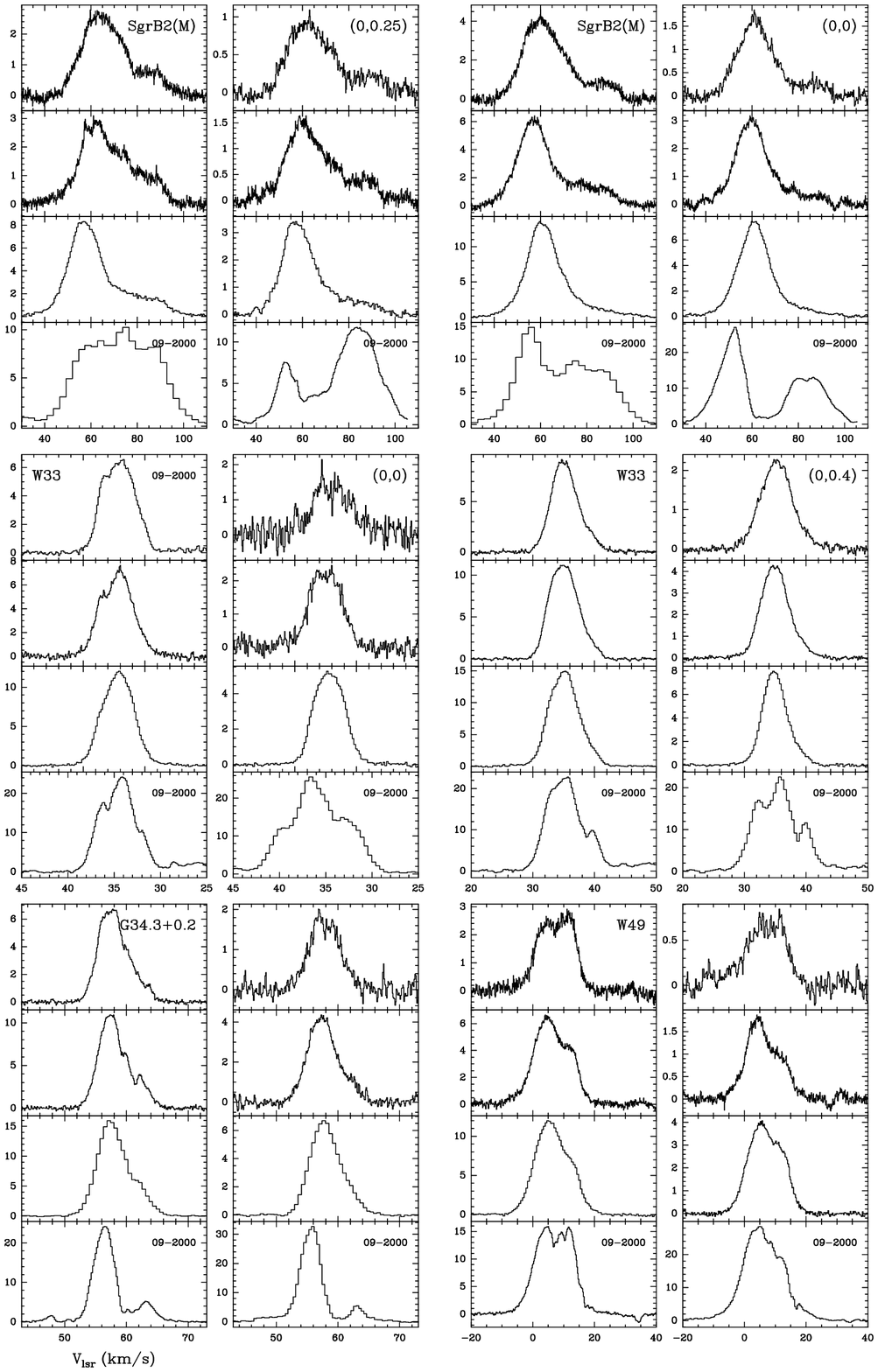}}
 \caption{Continued.}
\end{figure*}

\addtocounter{figure}{-1}

\begin{figure*}
 \resizebox{14.5cm}{!}{\includegraphics{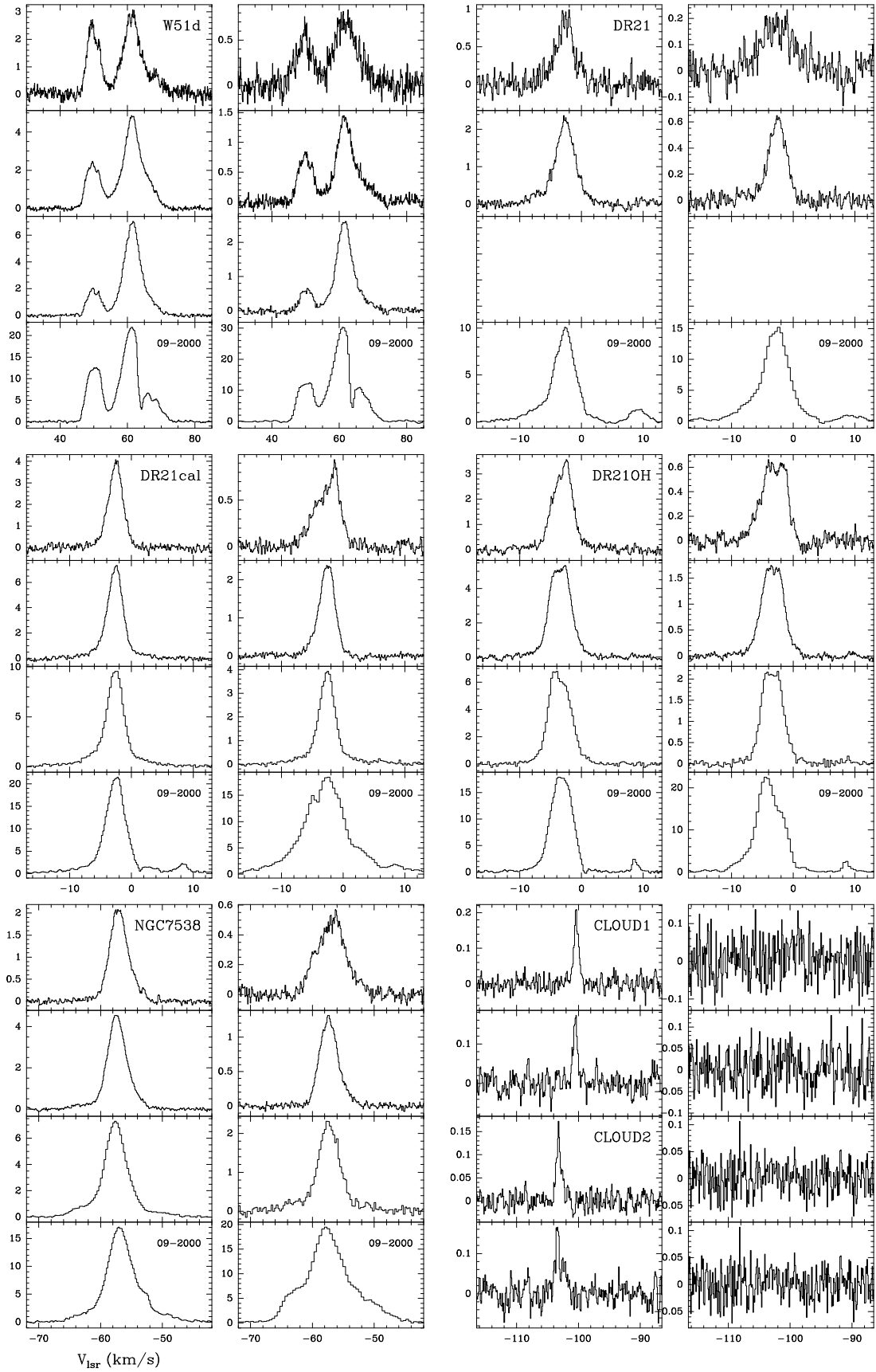}}
 \caption{Continued. Towards DDT94~Cloud 1 and 2 only the C$^{18}$O and 
C$^{17}$O $J$=1--0 and 2--1 lines have been observed.}
\end{figure*}

\end{document}